\documentclass[twocolumn,trackchanges]{aastex7}

\usepackage{graphicx}
\usepackage{multirow}
\usepackage{amsmath,amssymb,amsfonts}
\usepackage{amsthm}
\usepackage{mathrsfs}
\usepackage{xcolor}
\usepackage{booktabs}
\usepackage{rotating}

\theoremstyle{plain}

\theoremstyle{definition}

\theoremstyle{remark}

\colorlet{trackchange}{blue} 
\definecolor{darkgreen}{RGB}{0, 150, 0}
\begin{document}

\title[Explainable AI for Solar Flare Prediction]{Explainable AI for Solar Flare Prediction: Quantitative Magnetic Field Analysis of Model-Focused Regions}

\author[0009-0006-5180-8052]{Z. Zheng}
\affiliation{School of Astronomy and Space Science, Nanjing University, Nanjing 210023, China}
\affiliation{Key Laboratory of Modern Astronomy and Astrophysics, Ministry of Education, Nanjing 210023, China}
\email{zz\_nju@smail.nju.edu.cn}

\author[0000-0002-9264-6698]{Q. Hao}
\email{haoqi@nju.edu.cn}
\affiliation{School of Astronomy and Space Science, Nanjing University, Nanjing 210023, China}
\affiliation{Key Laboratory of Modern Astronomy and Astrophysics, Ministry of Education, Nanjing 210023, China}

\author[0000-0001-7693-4908]{C. Li}
\email{lic@nju.edu.cn}
\affiliation{School of Astronomy and Space Science, Nanjing University, Nanjing 210023, China}
\affiliation{Key Laboratory of Modern Astronomy and Astrophysics, Ministry of Education, Nanjing 210023, China}
\affiliation{Institute of Science and Technology for Deep Space Exploration, Suzhou Campus, Nanjing University, Suzhou 215163, China}

\author{P. F. Chen}
\affiliation{School of Astronomy and Space Science, Nanjing University, Nanjing 210023, China}
\affiliation{Key Laboratory of Modern Astronomy and Astrophysics, Ministry of Education, Nanjing 210023, China}
\email{chenpf@nju.edu.cn}

\author{J. R. Hu}
\affiliation{School of Astronomy and Space Science, Nanjing University, Nanjing 210023, China}
\affiliation{Key Laboratory of Modern Astronomy and Astrophysics, Ministry of Education, Nanjing 210023, China}
\email{jiaruihu@smail.nju.edu.cn}

\author[0000-0002-4978-4972]{M.D. Ding}
\affiliation{School of Astronomy and Space Science, Nanjing University, Nanjing 210023, China}
\affiliation{Key Laboratory of Modern Astronomy and Astrophysics, Ministry of Education, Nanjing 210023, China}
\email{dmd@nju.edu.cn}

\author{C. Fang}
\affiliation{School of Astronomy and Space Science, Nanjing University, Nanjing 210023, China}
\affiliation{Key Laboratory of Modern Astronomy and Astrophysics, Ministry of Education, Nanjing 210023, China}
\email{fangc@nju.edu.cn}

\correspondingauthor{Q. Hao, C. Li}
\email{haoqi@nju.edu.cn, lic@nju.edu.cn}

\begin{abstract}
Solar flares are intense energy release events in the solar atmosphere that may pose significant space weather hazards, which makes developing reliable prediction models essential. Although deep learning methods, particularly convolutional neural networks (CNNs), demonstrate strong predictive performance when using solar magnetograms, their scientific credibility is undermined by a lack of physical interpretability. Explainable artificial intelligence (XAI) offers a potential solution. However, current XAI studies in solar flare prediction are largely qualitative and lack systematic, theory-based, quantitative validation. We present a quantitative XAI framework that can decipher the physical basis of CNN-based solar flare prediction models. Using gradient-weighted class activation mapping (Grad-CAM), we identify model-focused regions (MFRs) in solar magnetograms. Then, we perform two key analyses to evaluate the predictive capability of magnetic parameters derived from MFRs and to quantitatively characterize their magnetic complexity. Our results reveal a strong physical correlation between MFRs and flare occurrence. Specifically, magnetic features extracted from MFRs demonstrate high predictive power for flares. Flare-producing active regions are characterized by magnetically complex configurations that are dominated by a single polarity rather than by balanced or purely unipolar structures. This finding is consistent with established physical theories of magnetic systems prone to flares. Our results suggest that CNNs can learn physically meaningful representations when trained on large-scale observations. Integrating XAI with quantitative magnetic field analysis improves the physical interpretability of deep learning-based flare prediction models, making them useful tools for prediction and modeling investigation in solar physics.
\end{abstract}

\section{Introduction}\label{sec_introduction}

Solar flares are among the most energetic phenomena in the solar atmosphere, occurring when accumulated magnetic energy is suddenly released \citep{Shibata2011}. These explosive events can accelerate charged particles and heat the surrounding plasma to tens of million degrees \citep{Fletcher2011}, potentially impacting the upper atmosphere of Earth and causing hazardous space weather effects \citep{Chen2011}. Therefore, reliable solar flare prediction models are therefore critical for space weather forecasting and early warning systems. Solar flare forecasting relies on observational data from space- and ground-based instruments, employing two main approaches: (1) physics-based models, which are grounded in theoretical understanding of solar processes \citep{Shibata2011,Toriumi2019}, and (2) knowledge-based techniques, which employ empirical and statistical methods to identify patterns in historical data \citep{Schrijver2007,Kusano2020}. In recent decades, machine learning has increasingly been applied to solar flare predictions \citep{asensio2023review,Huang2024}. Researchers have employed various approaches based on different data sources. One such approach uses space-weather HMI active region patches (SHARP) magnetic field parameters \citep{bobra2014sharp}, which are derived from the observation by the Helioseismic and Magnetic Imager (HMI; \citealt{hmi2012}) on board the Solar Dynamics Observatory (SDO; \citealt{sdo2012}) and applied to models such as support vector machines (SVM, \citealt{bobra2015solar}), random forest (RF, \citealt{liu2017predicting}), long short-term memory (LSTM) networks \citep{liu2019predicting}, and Transformers \citep{abduallah2023operational}. Another approach involves applying convolutional neural networks (CNNs) directly to magnetogram images \citep{huang2018deep,sun2022solar}. A third approach combines magnetic parameters and magnetogram images using ensemble learning or model fusion methods to further enhance prediction performance \citep{tang2021prediction,xu2025prediction}.

Despite their reasonably high success rate, deep learning models are often criticized as ``black boxes'' due to the opacity of decision-making processes \citep{asensio2023review}. This lack of interpretability poses two challenges. At the engineering level, it is difficult to assess the reliability of models for operational deployment, which limits their application in forecasting systems \citep{Huang2024}. At the scientific level, it is unclear whether the models have learned genuine physical mechanisms or merely exploit statistical correlations. This constrains the potential of artificial intelligence as an effective tool for scientific discovery. To address these challenges, explainable artificial intelligence (XAI) techniques have emerged \citep{Arrieta2020}. XAI encompasses two complementary research pathways. First, physics-informed neural networks (PINNs) embed physical laws directly into network architectures or loss functions, ensuring that the predictions adhere to the known physical constraints \citep{raissi2019physics}. These approaches have been applied to solar wind prediction \citep{costa2024Leveraging} and magnetic field inversion \citep{jarolim2025pinnme}. Second, interpretability techniques analyze trained models to extract patterns and features learned from data, including feature importance analysis, attention mechanism visualization, and gradient-based attribution methods \citep{samek2021explaining}.

As deep learning continues to advance in the fields of solar and astrophysics, XAI is becoming increasingly important for understanding the decision-making processes of models and evaluating their physical consistency \citep{lieu2025comprehensive, asensio2023review}. With feature-based machine learning models, the SHARP dataset has been widely used for flare prediction research \citep{bobra2015solar, liu2017predicting, zhang2022,Li2025,Newman2025ApJS}. Further analysis of feature importance based on feature-based models has revealed key physical quantities closely associated with flare occurrence, providing critical physical clues for understanding flare initiation and triggering mechanisms. Meanwhile, image-based XAI methods offer astronomers, as researchers in an observation-based science, a more direct and effective means of analyzing the spatial attention characteristics of models. \citet{huang2018deep}, \citet{yi2021solar}, and \citet{sun2022solar} applied image attribution methods to identify the most salient regions of images that contribute to CNN-based solar flare predictions. We refer to these regions as model-focused regions (MFRs). Their findings reveal that, during flare prediction, these MFRs tend to concentrate on regions near the polarity inversion line (PIL) and areas with highly complex magnetic structures. While these studies provide valuable qualitative insights, a systematic quantitative analysis of MFRs is lacking. Therefore, we conducted a quantitative magnetic field analysis of MFRs, demonstrating that CNNs can extract meaningful physical patterns from large-scale observations and capture magnetic properties relevant to solar flare activity.

\begin{figure*}[ht!]
    \centering
    \includegraphics[width=1.\linewidth]{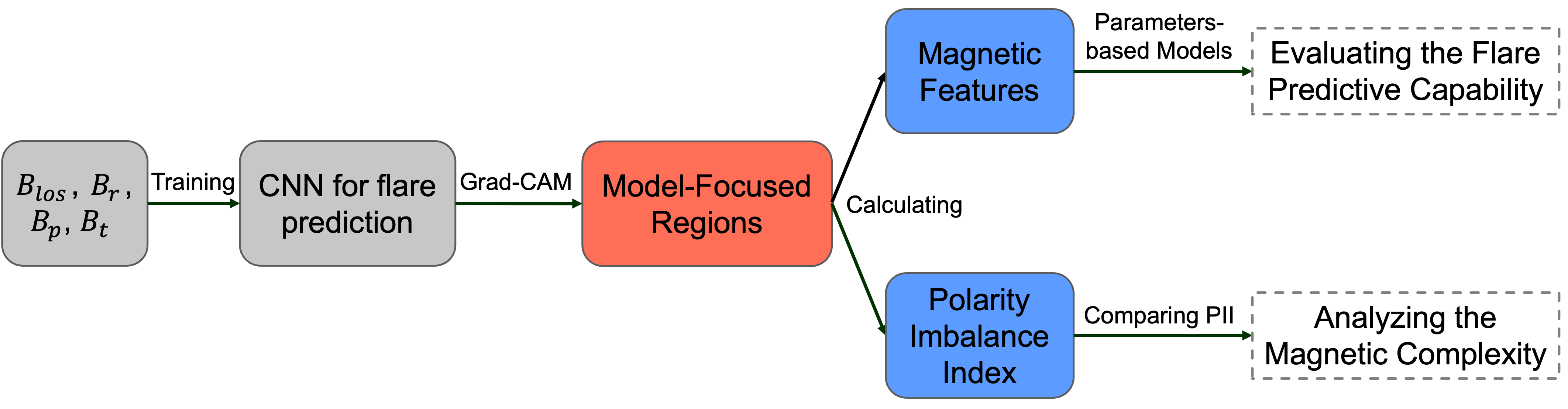}
    \caption{The flowchart of this study. It involves: (1) training a CNN-based flare prediction model using vector magnetic field data from solar active regions; (2) identifying MFRs via Grad-CAM; and (3) evaluating their physical significance through two quantitative experiments: comparing the predictive capability of parameter-based models using features extracted from MFR, SHARP, and PIL masks, and analyzing magnetic complexity via the polarity imbalance index.
    \label{fig_flowchart}}
\end{figure*}

Our aim was to quantitatively evaluate the scientific credibility of MFRs. First, we trained a CNN-based flare-prediction model using vector magnetic field data from the solar active regions. Then, we applied the gradient-weighted class activation mapping (Grad-CAM, \citealt{selvaraju2017grad}) to identify MFRs. To evaluate the physical significance of the MFRs identified by the CNN, we designed two quantitative experiments. The first experiment evaluated the flare prediction capability of sixteen magnetic-field parameters, derived from the magnetic fields extracted from the MFRs. The second experiment examined the magnetic complexity of the MFRs by focusing on the spatial distribution of positive and negative polarities within these regions in line-of-sight magnetogram. Figure~\ref{fig_flowchart} illustrates the overall workflow of our experiments. 

\section{Data}\label{sec_data}

This study employs comprising 2699 HMI active region patches (HARPs), accounting for around half of all HARPs observed between May 1, 2010, and September 30, 2022. To minimise the impact of projection effects, we only retained  active regions located within $\pm70^\circ$ of the central meridian. Furthermore, to facilitate subsequent labeling using GOES flare events, we only retained HARP IDs that were uniquely associated with a single NOAA active region number. Flare labels were assigned according to the operational scheme of \citet{ahmed2013solar}: a sample was labeled as positive if the corresponding active region produced a C-class or stronger flare within 24 hours after the observation time; otherwise, it was labeled as negative. Flare event information was obtained from the GOES-based catalog compiled by \citet{abduallah2023operational}, available at \url{https://nature.njit.edu/solardb/solarflarenet}.

In solar flare prediction, flaring events (positive samples) are typically much less frequent than non-flaring events (negative samples). This discrepancy may cause the trained model to favor the majority class, leading to overly optimistic performance estimates. To address this class imbalance, we retained all positive samples and downsampled the negative samples, as the original positive-to-negative ratio was approximately $1:20$. Specifically, only one negative sample was retained for each active region on each day. After these procedures and removing samples containing `NaN' values, the resulting dataset had an approximately balanced class distribution (around $1:1$) while preserving broad temporal and active region coverage.

The input to the CNN-based flare prediction model consisted of SHARP vector magnetic field data, including the line-of-sight magnetograms, the radial, poloidal, and toroidal components (corresponding to $B_{\mathrm{los}}$, $B_r$, $B_p$, and $B_t$, respectively). To satisfy the fixed-size input requirement of the CNN, we developed a standardized preprocessing pipeline for the raw magnetic field data. For each SHARP sample, we zero-padded the four magnetic field maps, $B_{\mathrm{los}}$, $B_r$, $B_p$, and $B_t$ to square matrices of size $\max(w,h) \times \max(w,h)$, where $w$ and $h$ denote the original width and height of the active-region patch, respectively. The padded maps were then resized to $512 \times 512$ pixels using bilinear interpolation. Finally, we concatenated the four processed magnetic field maps along the channel dimension, yielding a standardized input tensor with the shape of $(4, 512, 512)$. This resizing procedure preserved the full active-region context and aspect ratio after zero-padding.

To evaluate the predictive capability of flare-related parameters derived from MFR, we compared them with the same set of parameters extracted from two physically motivated reference regions: the SHARP mask and the PIL mask. The SHARP mask is defined in \citet{bobra2014sharp}, and the PIL maskis defined in \citet{ran2022solar}. Specifically, for each mask type, we computed sixteen magnetic field parameters: \texttt{TOTUSJH}, \allowbreak \texttt{TOTPOT}, \allowbreak \texttt{TOTUSJZ}, \allowbreak \texttt{ABSNJZH}, \allowbreak \texttt{SAVNCPP}, \allowbreak \texttt{USFLUX}, \allowbreak \texttt{MEANPOT}, \allowbreak \texttt{R\_VALUE}, \allowbreak \texttt{MEANSHR}, \allowbreak \texttt{MEANGAM}, \allowbreak \texttt{MEANGBT}, \allowbreak \texttt{MEANGBZ}, \allowbreak \texttt{MEANGBH}, \allowbreak \texttt{MEANJZD}, \allowbreak \texttt{MEANJZH}, and \allowbreak \texttt{MEANALP}. These parameters were computed based on the publicly available implementations of \citet{bobra2021github} and \citet{ran2022solar}. We then used the resulting parameter sets extracted from the MFRs, SHARP mask, and PIL mask as inputs to the same parameter-based prediction models. This allowed us to assess the predictive utility of the MFRs relative to these physically motivated reference regions. The definitions and generation procedures of the three masks are described in Section~\ref{subsec_masks}.

\subsection{Data Splitting}\label{sec_data_splitting}

\begin{figure}[ht!]
    \centering
    \includegraphics[width=\linewidth]{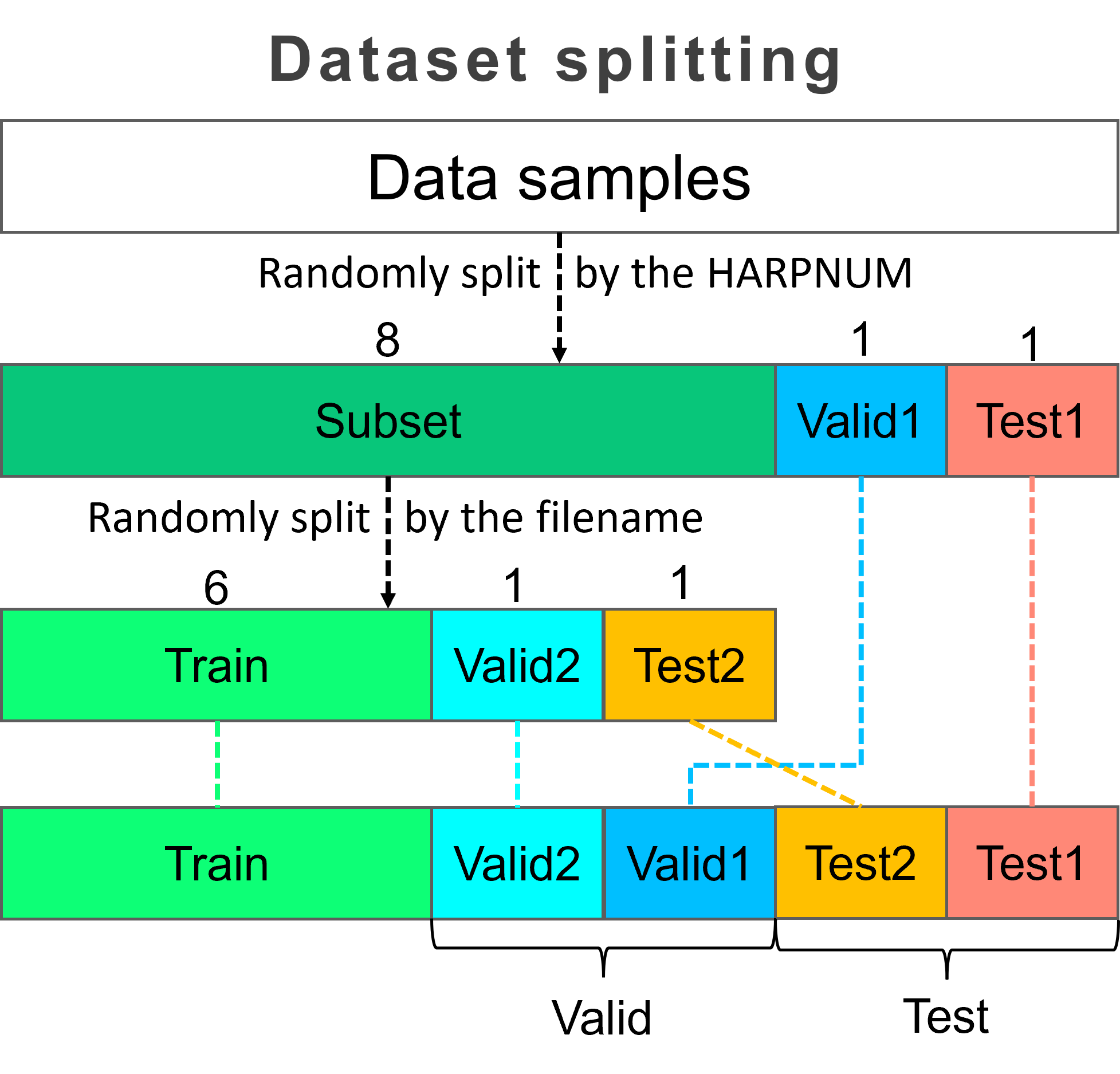}
    \caption{The flowchart of the data splitting.
    \label{fig_datasplit}}
\end{figure}

Data splitting can substantially affect the evaluation of prediction models. Common strategies include random splitting, chronological splitting, flare-level-based splitting, and active-region-based splitting. Among these, splitting by active region is particularly important because it prevents samples from the same active region from appearing in both the training and test sets. Ignoring this constraint may introduce information leakage and lead to overly optimistic estimates of model generalization performance \citep{sun2022predicting}. To examine this effect, we adopted a two-stage data splitting strategy, as illustrated in Figure~\ref{fig_datasplit}. 

First, the dataset was split by active region to construct \textit{Test1} and \textit{Valid1}, ensuring that these two subsets did not active-region overlap with the training data. Then, the remaining samples were then randomly divided into the training set, \textit{Test2}, and \textit{Valid2} using the filename. Thus, \textit{Test2} and \textit{Valid2} have active region overlap with the training dataset. Finally, the validation and test sets were obtained by combining \textit{Valid1} with \textit{Valid2} and \textit{Test1} with \textit{Test2}, respectively. Overall, the training, validation, and test sets followed an approximate ratio of 3:1:1. Each of \textit{Valid1}, \textit{Valid2}, \textit{Test1}, and \textit{Test2} accounts for about $10\%$ of the total dataset. We assessed the influence of active-region information leakage on performance estimation by comparing model performance on \textit{Test1}, which had no active-region overlap with the training set, and \textit{Test2}, which allowed active-region overlap. The detailed sample counts for each subset are summarized in Table~\ref{tab_dataset}.

\begin{table}[!ht]
\centering
\caption{Summary of dataset splits and class distributions.}
\label{tab_dataset}
\begin{tabular}{ccccc}
\hline
\textbf{Dataset} & \textbf{Subset} & \textbf{AR Overlap} & \textbf{Negative} & \textbf{Positive} \\
\hline
Training &   &  & 9416 & 9127 \\
\hline
\multirow{2}{*}{Validation}
 & Valid1 & No  & 1454 & 976  \\
 & Valid2 & Yes & 1552 & 1539 \\
\hline
\multirow{2}{*}{Test}
 & Test1  & No  & 1587 & 1507 \\
 & Test2  & Yes & 1539 & 1552 \\
\hline
\end{tabular}
\end{table}

Both the CNN and the parameter-based models, including SVM, RF, and Fully Connected Neural Network (FCNN, \citealt{rumelhart1986learning}), were trained using the training set. The validation set was used for model selection, and the test set was reserved for final performance evaluation. Since SVM and RF do not require an independent validation stage for iterative training, the training and validation sets were combined when fitting these two models. Their performance was then evaluated using the test set.

\subsection{Generation of MFRs and Reference Masks}\label{subsec_masks}

\begin{figure*}[ht!]
    \centering
    \includegraphics[width=\linewidth]{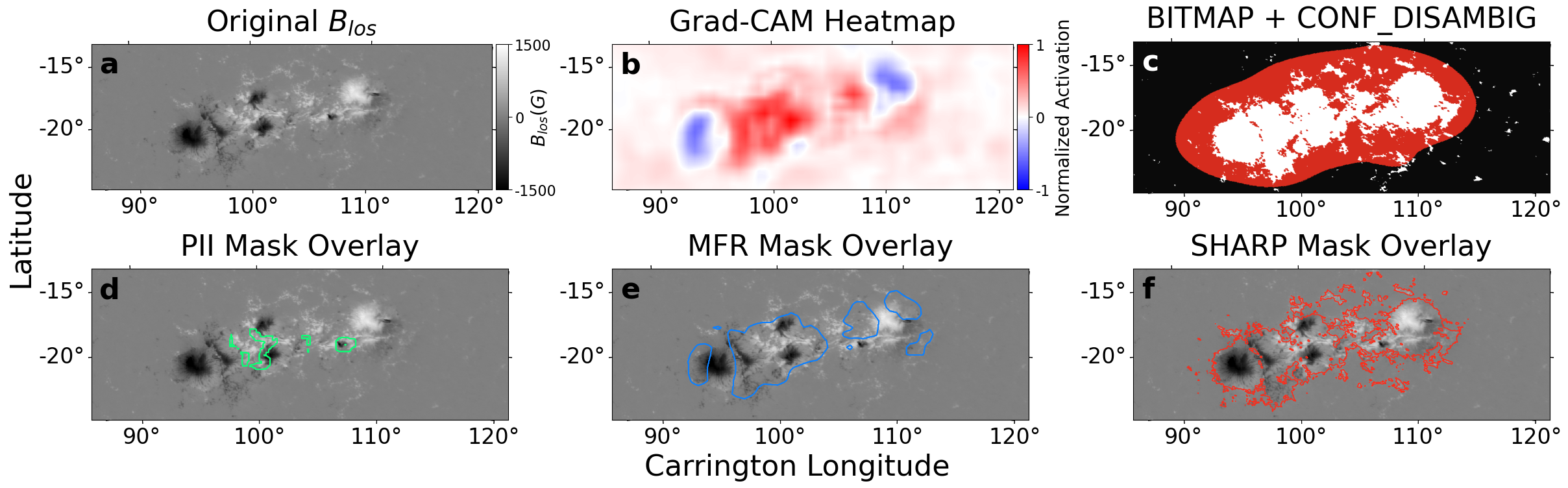}
    \caption{An example (HARP 5983 on 30 September 2015 at 03:00:00 UT) illustrating different mask types used in the analysis. (a) The line-of-sight magnetogram. (b) The Grad-CAM heatmap. (c) The overlapping region where $\text{BITMAP} \geq 30$ (red) and $\text{CONF\_DISAMBIG} \geq 70$ (white). (d) The PIL mask contours (green). (e) The MFR mask contours (blue). (f) The SHARP mask contours (red). 
    \label{fig_masks}}
\end{figure*}

We first trained a CNN model to predict solar flare eruptions using vector magnetic field data from solar active regions. After confirming its predictive performance, we applied the Grad-CAM (\citealt{selvaraju2017grad}) to identify the image regions that contributed most to the CNN predictions, defined here as MFRs. Grad-CAM has been widely used in solar flare prediction to determine which regions in the input data are more strongly associated with flare occurrence \citep{yi2021solar, sun2022predicting, panos2023identifying}. Following the recommendation in \citet{selvaraju2017grad}, we selected the last convolutional layer as the target for Grad-CAM. This layer typically captures high-level semantic features while retaining sufficient spatial resolution, making it well-suited for localizing regions in the input that influence the model's prediction. In our model, this layer corresponded to the second convolutional layer of the third residual block of the CNN introduced in Section~\ref{subsec_cnn}.

Grad-CAM produces a class activation heatmap by computing a weighted sum of the feature maps $A^k$ from the target convolutional layer. The weights $\alpha^k$ are obtained by global average pooling of the gradients of the output score $y$ with respect to each feature map:

\begin{equation}
\text{Grad-CAM heatmap} = \sum_{k=1}^K \alpha^k A^k.
\end{equation}

Unlike the original Grad-CAM implementation \citep{selvaraju2017grad}, which uses the rectified linear unit (ReLU) activation function \citep{nari2010relu} to retain only positive activations, we preserved both positive and negative values in order to capture regions that may either promote or suppress flare activity. Both of these are physically meaningful in our context. To convert the heatmap into a binary mask, we apply OTSU thresholding \citep{otsu1979ieee} to its absolute value. Unlike fixed thresholds (e.g., 0.5), which are applied after normalization and do not necessarily correspond to the most relevant or informative regions, OTSU provides a data-driven alternative that selects thresholds adaptively by minimizing intra-class variance. This yields more meaningful region separation. The final MFR mask is defined as:

\begin{equation}
\text{MFR mask} = \text{OTSU}(abs(\text{Grad-CAM heatmap})).
\end{equation}

To compare with the MFR mask, we considered two physics-based regions: the SHARP mask and the PIL mask. The SHARP mask delineates the regions used to calculate SHARP magnetic parameters. It is constructed based on two HMI data products: \text{BITMAP}, which encodes the active region boundary, and \text{CONF\_DISAMBIG}, which indicates the confidence in azimuthal disambiguation of the transverse magnetic field. The SHARP mask is defined as the intersection of regions where $\text{BITMAP} \geq 30$ and $\text{CONF\_DISAMBIG} \geq 70$ \citep{bobra2021github}.

PILs are widely regarded as key sites for solar flare initiation and have been extensively studied in the context of flare-productive active regions \citep{falconer2002solar, schrijver2007solar, wang2019parameters}. To extract the PIL mask, we adopt the implementation provided by \citet{ran2022solar}, available at \url{https://github.com/RanHao1999/Flare_SHARP.git}. This method identifies regions with strong opposite magnetic polarities ($B_r > 200$ G and $B_r < -200$ G), applies Gaussian dilation (with $\sigma=10$) to both polarities, and marks the overlapping regions as the PIL. To further refine the mask, we apply the Density-Based Spatial Clustering of Applications with Noise (DBSCAN, \citealt{Schubert2017}) algorithm to eliminate isolated small regions and retain only the five largest contiguous areas, forming the final PIL mask.

Figure~\ref{fig_masks} illustrates an example of the data and masks used in our analysis. Figures~\ref{fig_masks}(a)--(c) show the input magnetogram, the corresponding Grad-CAM heatmap, and the overlapping region defined by the \text{BITMAP} and \text{CONF\_DISAMBIG} thresholds, respectively. Figures~\ref{fig_masks}(d)--(f) present the masks corresponding to the PIL, MFR, and SHARP regions, respectively. For each masked region, we computed the sixteen magnetic features mentioned above, resulting in three distinct parameter sets corresponding to the MFR, SHARP, and PIL masks.

\section{Methods}\label{sec_methods}

\subsection{Convolutional Neural Network}\label{subsec_cnn}

\begin{figure*}[ht!]
    \centering
    \includegraphics[width=\linewidth]{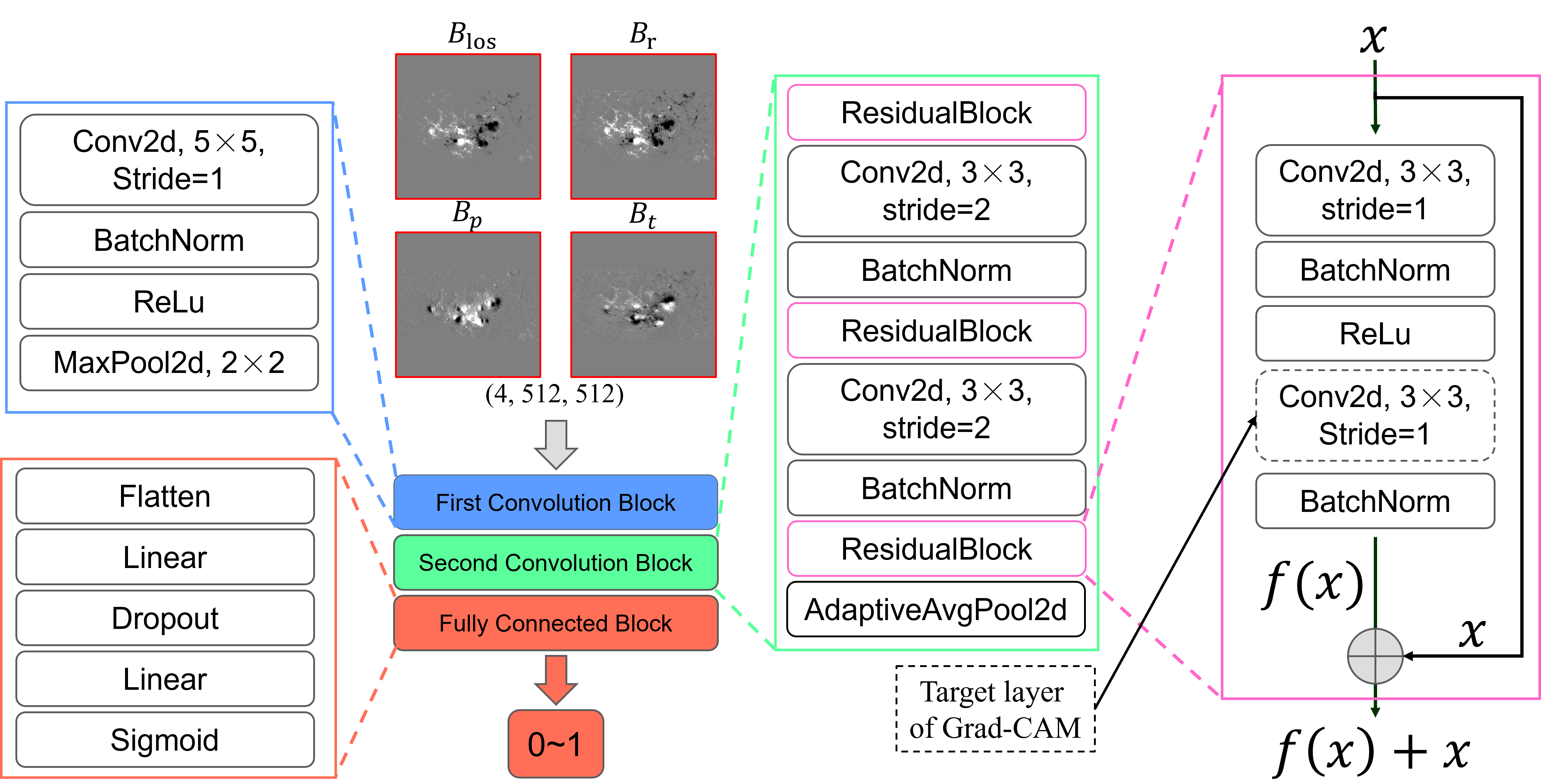}
    \caption{Schematic representation of the CNN model used in this work. Four-channel magnetic field images ($B_{\mathrm{los}}, B_r, B_p$, and $B_t$) with an input size $(4,512,512)$ are processed through convolutional and residual blocks, followed by global average pooling and sigmoid classification to predict the probability of flare occurrence within 24 hours. The final convolutional layer, which is used as the target layer for Grad-CAM, is highlighted.
    \label{fig_cnnmodel}}
\end{figure*}

We established a baseline CNN model with robust performance. This model is a standard variant of the ResNet architecture \citep{he2016deep}. After training the model to meet the desired performance metrics, we used the Grad-CAM method to visualize the MFRs. The model consists of three primary components:

\textbf{First Convolutional Layer}: The model takes a four-channel image as input, composed of $B_{\mathrm{los}}, B_r, B_p$, and $B_t$, which represent different components of the photospheric magnetic field. Using $5\times5$ convolution kernels and max pooling operations, this layer extracts primary features. 

\textbf{Second Convolutional Layers}: Feature extraction is performed using a series of convolutional layers interleaved with residual blocks. The convolution kernels gradually expand the number of feature channels while reducing spatial resolution through strided convolutions. Using strided convolutions instead of traditional pooling allows the network to adaptively learn and retain salient features while reducing spatial resolution. Residual connections allow the network to preserve low-level features and improve training stability.

\textbf{Fully Connected Layer}: A global average pooling (GAP) layer \citep{lin2014network} is applied before the output layer to summarize spatial information into a compact representation. A final dense layer with a sigmoid activation function outputs a probability between 0 and 1, and a threshold of 0.5 is used for binary classification. Dropout \citep{srivastava2014dropout} is included to reduce overfitting.

Figure~\ref{fig_cnnmodel} shows an overview of the model architecture, and Table~\ref{tab_cnnmodel} lists the corresponding layer configurations.

\begin{table}[!ht]
\centering
\caption{CNN architecture in this work}\label{tab_cnnmodel}
\begin{tabular}{ll}
\hline
\textbf{Layer / Operation} & \textbf{Shape of Output} \\
\hline
Input & (4, 512, 512) \\
Conv2d(5$\times$5, 64, padding=2) & (64, 512, 512) \\
MaxPool2d(2) & (64, 256, 256) \\
ResidualBlock(64) & (64, 256, 256) \\
Conv2d(3$\times$3, 128, stride=2) & (128, 128, 128) \\
ResidualBlock(128) & (128, 128, 128) \\
Conv2d(3$\times$3, 256, stride=2) & (256, 64, 64) \\
ResidualBlock(256) & (256, 64, 64) \\
AdaptiveAvgPool2d(1) & (256, 1, 1) \\
Flatten & (256) \\
Linear(256$\rightarrow$128) & (128) \\
Linear(128$\rightarrow$1) & (1) \\
Sigmoid & (1) \\
\hline
\end{tabular}
\end{table}

Before generating MFRs, we evaluated the performance of the trained CNN on the test set to ensure reasonable flare prediction. As shown in Table~\ref{tab_cnnperform}, the CNN achieved a precision of 0.860, a recall of 0.890, an F1 score of 0.875, and a TSS of 0.748 on the combined test set. These results suggest that the trained CNN captured the flare-relevant patterns from the magnetic field data, providing a solid foundation for the subsequent MFR analysis.

In addition, the two-stage data splitting strategy enabled us to compare the CNN performance on \textit{Test1} and \textit{Test2}. The small differences in TSS and other metrics between these two subsets suggest that active-region overlap may introduce optimistic bias into model evaluation to some degree. However, a systematic investigation of the impact of active-region information leakage is beyond the scope of this work. Therefore, when comparing different parameter groups, we evaluated the performance of all parameter-based models on the combined \textit{Test} set to maintain a consistent evaluation protocol.

\begin{table}[!ht]
    \centering
    \caption{CNN model evaluation metrics for different datasets \label{tab_cnnperform}}
    \begin{tabular}{ccccc}
    \hline
    \textbf{Dataset} & \textbf{Precision} & \textbf{Recall} & \textbf{F1 Score} & \textbf{TSS} \\
    \hline
    \textit{Test}  & 0.860 & 0.890 & 0.875 & 0.748 \\
    \textit{Test1} & 0.856 & 0.873 & 0.865 & 0.734 \\
    \textit{Test2} & 0.864 & 0.907 & 0.885 & 0.762 \\
    \hline
    \end{tabular}
\end{table}

\subsection{Parameters-based models}\label{subsec_parameters-based models}

We employed three heterogeneous machine learning classification models to systematically and quantitatively evaluate the effectiveness of three sets of parameter data in solar flare prediction tasks: RF, FCNN, and SVM. These models are based on ensemble learning, deep learning, and statistical learning theories, respectively, and have been used in various studies on parameter-based solar flare prediction studies \citep{bobra2015solar, nishizuka2018prediction}.

RF is an ensemble learning method that constructs multiple decision trees and combines their predictions for classification \citep{breiman2001random}. SVM \citep{Hearst1998} is a classification algorithm based on statistical learning theory that constructs hyperplanes in high-dimensional space to achieve classification \citep{cortes1995support}. In this study, both classifiers were implemented using the scikit-learn library. Specifically, the RF model was built using the \texttt{RandomForestClassifier} class from the \texttt{sklearn.ensemble} module, and the SVM was implemented using the \texttt{SVC} class from the \texttt{sklearn.svm} module. FCNN is a feedforward neural network composed of multiple fully connected layers that are trained via backpropagation \citep{rumelhart1986learning}. Our FCNN model has two hidden layers, each of which is followed by a LeakyReLU activation function and Dropout regularization. The output layer uses a Sigmoid function for binary classification. The model structure is shown in Table~\ref{tab_fcnnmodel}.

The parameter-based model used sixteen magnetic features (\texttt{TOTUSJH}, \allowbreak \texttt{TOTPOT}, \allowbreak \texttt{TOTUSJZ}, \allowbreak \texttt{ABSNJZH}, \allowbreak \texttt{SAVNCPP}, \allowbreak \texttt{USFLUX}, \allowbreak \texttt{MEANPOT}, \allowbreak \texttt{R\_VALUE}, \allowbreak \texttt{MEANSHR}, \allowbreak \texttt{MEANGAM}, \allowbreak \texttt{MEANGBT}, \allowbreak \texttt{MEANGBZ}, \allowbreak \texttt{MEANGBH}, \allowbreak \texttt{MEANJZD}, \allowbreak \texttt{MEANJZH}, and \allowbreak \texttt{MEANALP}) calculated under different masks, i.e., SHARP, MFR, and PIL. The definitions of the sixteen magnetic features are the same as those in \citealt{bobra2014sharp} (see Table 3 in their paper for details).

\begin{table}[htbp]
\centering
\caption{FCNN architecture in this work}
\label{tab_fcnnmodel}
\begin{tabular}{ll}
\toprule
\textbf{Layer/Operation} & \textbf{Shape of Output} \\
\midrule
Input & (16) \\
Linear (16 $\rightarrow$ 128) & (128) \\
LeakyReLU ($\textit{leaky\_alpha}$) & (128) \\
Dropout ($\textit{dropout\_rate}$) & (128) \\
Linear (128 $\rightarrow$ 64) & (64) \\
LeakyReLU ($\textit{leaky\_alpha}$) & (64) \\
Dropout ($\textit{dropout\_rate}$) & (64) \\
Linear (64 $\rightarrow$ 1) & (1) \\
Sigmoid & (1) \\
\bottomrule
\end{tabular}
\end{table}

To systematically evaluate the performance of the three parameter sets across the three models, we performed hyperparameter tuning on each model. Based on considerations of model initialization, classification methodology, and robustness, we adjusted the following hyperparameters: the random seed, which can affect the model initialization; the number of trees ($n\_estimators$) and maximum tree depth ($max\_depth$) for RF; LeakyReLU $\alpha$ value ($leaky\_alpha$) and dropout rate ($dropout\_rate$) for FCNN; the kernel function ($kernel$) and regularization parameter ($C$) for SVM. The specific hyperparameter selections are presented in Table~\ref{tab_hyperparam}, while the other parameters were kept at their default values. 

\begin{table}[htbp]
\centering
\caption{Hyperparameter ranges for different parameter-based models}
\label{tab_hyperparam}
\begin{tabular}{lll}
\toprule
\textbf{Model} & \textbf{Hyperparameters} & \textbf{Range} \\
\midrule
\textit{(All Models)} & random seed & \{2, 4, 8, 16, 32\} \\
\midrule
\multirow{2}{*}{RF} 
  & n\_estimators & \{50, 100, 1000\} \\
  & max\_depth    & \{5, 10, 20, None\} \\
\midrule
\multirow{2}{*}{FCNN} 
  & leaky\_alpha  & \{0, 0.01, 0.1\} \\
  & dropout\_rate & \{0.1, 0.3, 0.5, 0.7\} \\
\midrule
\multirow{2}{*}{SVM} 
  & kernel        & \{rbf, poly, sigmoid\} \\
  & C             & \{0.1, 1, 100, 1000\} \\
\bottomrule
\end{tabular}
\end{table}

\section{Results}\label{sec_results}

\subsection{Predictive Capability Evaluation of MFR-derived parameters}\label{subsec_prediction capability}

The baseline CNN model was trained using the active region vector magnetic field (line-of-sight $B_{\mathrm{los}}$, radial $B_r$, and poloidal and toroidal components $B_p$ and $B_t$) from the SHARP dataset. We assigned flare labels based on the GOES flare catalogs. The model outputs a probabilistic score (0--1) representing the likelihood of a $\ge$ C-class flare occurring within the next 24 hours in the given active region. The CNN model achieved a true skill statistic (TSS) of 0.748 on the test set. This indicates that the model effectively discriminates between positive and negative samples rather than making random predictions. Thus, it is confirmed that the MFRs identified by the CNN serve as critical features for flare prediction.

We then applied Grad-CAM and OTSU thresholding algorithm to identify the MFRs (as shown in Figure~\ref{fig_masks}e). Unlike traditional Grad-CAM which only outputs positive activation maps, we retained the negative activations, which may correspond to regions with low correlation or inhibitory effects on flare production. We evaluated the predictive capability of sixteen magnetic features, commonly used for flare prediction in the SHARP dataset. These features were extracted from the MFRs. We compared these features side-by-side with the parameters derived from two alternative physics-based regions: the SHARP mask \citep{bobra2021github} and the PIL mask \citep{ran2022solar} (as shown in Figures~\ref{fig_masks}d and f). For a systematic evaluation, we employed three heterogeneous classification models: SVM \citep{Hearst1998}, RF \citep{Breiman2001}, and a FCNN. Each model was trained with 60 different hyperparameter configurations to ensure the robustness and generalization. 

\begin{table*}[ht!]
    \centering
    \caption{Mean evaluation metrics for different models and datasets}
    \label{tab_predictive_capability}
    \scriptsize
    \begin{tabular}{cc|cccccc}
    \hline
    Model & Dataset & Accuracy & Precision & Recall & Specificity & F1 & TSS \\
    \hline

    \multirow{3}{*}{FCNN}
    & MFR & 0.8549 & 0.8560 & \textbf{\textcolor{darkgreen}{0.9639}} & 0.5578 & 0.9067 & 0.5217 \\
    & SHARP
      & \textbf{\textcolor{darkgreen}{0.8572}}
      & \textbf{\textcolor{darkgreen}{0.8596}}
      & 0.9621
      & \textbf{\textcolor{darkgreen}{0.5715}}
      & \textbf{\textcolor{darkgreen}{0.9079}}
      & \textbf{\textcolor{darkgreen}{0.5336}} \\
    & PIL      & 0.8109 & 0.8267 & 0.9381 & 0.4640 & 0.8789 & 0.4022 \\

    \cmidrule{1-8}

    \multirow{3}{*}{RF}
    & MFR & 0.8600 & 0.8614 & \textbf{\textcolor{darkgreen}{0.9638}} & 0.5770 & 0.9097 & 0.5408 \\
    & SHARP
      & \textbf{\textcolor{darkgreen}{0.8621}}
      & \textbf{\textcolor{darkgreen}{0.8660}}
      & 0.9607
      & \textbf{\textcolor{darkgreen}{0.5937}}
      & \textbf{\textcolor{darkgreen}{0.9107}}
      & \textbf{\textcolor{darkgreen}{0.5543}} \\
    & PIL      & 0.8216 & 0.8344 & 0.9435 & 0.4896 & 0.8856 & 0.4331 \\

    \cmidrule{1-8}

    \multirow{3}{*}{SVM}
    & MFR & \textbf{\textcolor{darkgreen}{0.8012}} & \textbf{\textcolor{darkgreen}{0.8319}} & \textbf{\textcolor{darkgreen}{0.9113}} & \textbf{\textcolor{darkgreen}{0.5012}} & \textbf{\textcolor{darkgreen}{0.8690}} & \textbf{\textcolor{darkgreen}{0.4125}} \\
    & SHARP    & 0.7919 & 0.8249 & 0.9074 & 0.4771 & 0.8635 & 0.3844 \\
    & PIL      & 0.7743 & 0.8191 & 0.8867 & 0.4683 & 0.8512 & 0.3550 \\

    \hline
    \end{tabular}
    \begin{flushleft}
    \scriptsize
    Note: The evaluation metrics are defined as follows: \\
    Accuracy = $(TP + TN) / (TP + TN + FP + FN)$;
    Precision = $TP / (TP + FP)$; \\
    Recall = $TP / (TP + FN)$;
    Specificity = $TN / (TN + FP)$; \\
    F1 = $2 \times \text{Precision} \times \text{Recall} / (\text{Precision} + \text{Recall})$;
    TSS = $Recall + Specificity - 1$, \\
    where $TP, TN, FP, FN$ refer to true positives, true negatives, false positives, and false negatives, respectively.
    \end{flushleft}
\end{table*}

Table~\ref{tab_predictive_capability} presents the mean performance metrics of the three parameter sets (aligned for consistency) across the three heterogeneous models. The SHARP dataset is a suitable benchmark because of its extensive use in solar flare prediction and the physical analysis of flare mechanisms. As shown in the table, the performance of the MFR group is comparable to that of the SHARP group and slightly superior to the PIL group. This suggests that the physical parameters extracted from the MFR masks are as effective as those from the SHARP masks for our prediction task. Furthermore, since many active regions lack prominent PIL regions, we evaluated the performance of the two groups (MFR and SHARP), which is shown in Table~\ref{appendix_tab_paramperform} and reflected in the confusion matrix sections of the two- and three-group alignments. In the two-group alignment, the MFR group demonstrated comparable predictive capability to the SHARP group. These results confirm that the MFRs identified by the CNN model contain meaningful physical information.

\begin{table*}[ht!]
    \centering
    \caption{Overall performance of different datasets across models based on different parameters}
    \label{appendix_tab_paramperform}
    \scriptsize
    \renewcommand{\arraystretch}{2}
    \setlength{\tabcolsep}{6pt}
    \begin{tabular}{ccc|cccc|cccccc}
    \hline
    \multirow{2}{*}{Alignment} & \multirow{2}{*}{Model} & \multirow{2}{*}{Dataset} &
    \multicolumn{4}{c|}{Confusion Matrix} & \multicolumn{6}{c}{Evaluation Metrics} \\
    \cmidrule{4-7} \cmidrule{8-13}
     & & & TP & FP & TN & FN & Accuracy & Precision & Recall & Specificity & F1 & TSS \\
    \hline
    \multirow{9}{*}{3-Group}
     & \multirow{3}{*}{FCNN}
     & MFR & 2395 & 403 & 509 & 90 & 0.85 & 0.86 & \textbf{\textcolor{darkgreen}{0.96}} & 0.56 & 0.91 & 0.52 \\
     & & SHARP    & 2391 & 391 & 521 & 94  & \textbf{\textcolor{darkgreen}{0.86}} & \textbf{\textcolor{darkgreen}{0.86}} & 0.96 & \textbf{\textcolor{darkgreen}{0.57}} & \textbf{\textcolor{darkgreen}{0.91}} & \textbf{\textcolor{darkgreen}{0.53}} \\
     & & PIL      & 2331 & 489 & 423 & 154 & 0.81 & 0.83 & 0.94 & 0.46 & 0.88 & 0.40 \\
    \cmidrule{2-13}
     & \multirow{3}{*}{RF}
     & MFR & 2395 & 386 & 526 & 90 & 0.86 & 0.86 & \textbf{\textcolor{darkgreen}{0.96}} & 0.58 & 0.91 & 0.54 \\
     & & SHARP    & 2387 & 371 & 541 & 98  & \textbf{\textcolor{darkgreen}{0.86}} & \textbf{\textcolor{darkgreen}{0.87}} & 0.96 & \textbf{\textcolor{darkgreen}{0.59}} & \textbf{\textcolor{darkgreen}{0.91}} & \textbf{\textcolor{darkgreen}{0.55}} \\
     & & PIL      & 2345 & 466 & 447 & 140 & 0.82 & 0.83 & 0.94 & 0.49 & 0.89 & 0.43 \\
    \cmidrule{2-13}
     & \multirow{3}{*}{SVM}
     & MFR & 2265 & 455 & 457 & 220 & \textbf{\textcolor{darkgreen}{0.80}} & \textbf{\textcolor{darkgreen}{0.83}} & \textbf{\textcolor{darkgreen}{0.91}} & \textbf{\textcolor{darkgreen}{0.50}} & \textbf{\textcolor{darkgreen}{0.87}} & \textbf{\textcolor{darkgreen}{0.41}} \\
     & & SHARP    & 2255 & 477 & 435 & 230 & 0.79 & 0.82 & 0.91 & 0.48 & 0.86 & 0.38 \\
     & & PIL      & 2203 & 485 & 427 & 282 & 0.77 & 0.82 & 0.89 & 0.47 & 0.85 & 0.36 \\
    \hline
    \multirow{6}{*}{2-Group}
     & \multirow{2}{*}{FCNN}
     & MFR & 2766 & 615 & 2511 & 293 & 0.85 & 0.82 & 0.90 & 0.80 & 0.86 & 0.71 \\
     & & SHARP    & 2794 & 604 & 2522 & 265 & \textbf{\textcolor{darkgreen}{0.86}} & \textbf{\textcolor{darkgreen}{0.82}} & \textbf{\textcolor{darkgreen}{0.91}} & \textbf{\textcolor{darkgreen}{0.81}} & \textbf{\textcolor{darkgreen}{0.87}} & \textbf{\textcolor{darkgreen}{0.72}} \\
    \cmidrule{2-13}
     & \multirow{2}{*}{RF}
     & MFR & 2765 & 568 & 2558 & 294 & 0.86 & 0.83 & 0.90 & 0.82 & 0.87 & 0.72 \\
     & & SHARP    & 2774 & 539 & 2587 & 285 & \textbf{\textcolor{darkgreen}{0.87}} & \textbf{\textcolor{darkgreen}{0.84}} & \textbf{\textcolor{darkgreen}{0.91}} & \textbf{\textcolor{darkgreen}{0.83}} & \textbf{\textcolor{darkgreen}{0.87}} & \textbf{\textcolor{darkgreen}{0.73}} \\
    \cmidrule{2-13}
     & \multirow{2}{*}{SVM}
     & MFR & 2605 & 746 & 2380 & 454 & 0.81 & 0.78 & 0.85 & 0.76 & 0.81 & 0.61 \\
     & & SHARP    & 2647 & 742 & 2384 & 412 & \textbf{\textcolor{darkgreen}{0.81}} & \textbf{\textcolor{darkgreen}{0.78}} & \textbf{\textcolor{darkgreen}{0.87}} & \textbf{\textcolor{darkgreen}{0.76}} & \textbf{\textcolor{darkgreen}{0.82}} & \textbf{\textcolor{darkgreen}{0.63}} \\
    \hline
\end{tabular}
\begin{flushleft}
\scriptsize
Note: The 3-group alignment denotes samples for which all three mask-based parameter sets (MFR, SHARP, and PIL) are available. The 2-group alignment denotes samples aligned between the MFR and SHARP parameter sets because PIL masks are unavailable for some HARPs. All reported values are the mean performance for the corresponding dataset and model. Bold green values indicate the maximum value of each evaluation metric within each model group.
\end{flushleft}
\end{table*}

\subsection{Magnetic Complexity Analysis of MFRs}\label{subsec_magnetic complexity}

The complexity of solar active regions is a key indicator in solar physics and space weather prediction \citep{Toriumi2019}. Its determination involves studying the magnetic field structure, evolution, and energy accumulation in these regions to assess the potential for eruptive events, such as solar flares and coronal mass ejections (CMEs). The complexity of solar active regions is typically characterized by a variety of physical parameters, including magnetic topological complexity (\textit{e.g.}, PIL, shear angle, and non-potentiality), magnetic parameter complexity (\textit{e.g.}, total magnetic flux, total vertical current and magnetic helicity), morphological complexity (\textit{e.g.}, sunspot group structure, active region area and multipolar structure), and the temporal evolution complexity, including evolution of active regions and dynamic changes in magnetic fields (see a review \cite{Toriumi2019} and the references therein). 

\begin{figure*}[ht!]
    \centering
    \includegraphics[width=0.9\textwidth]{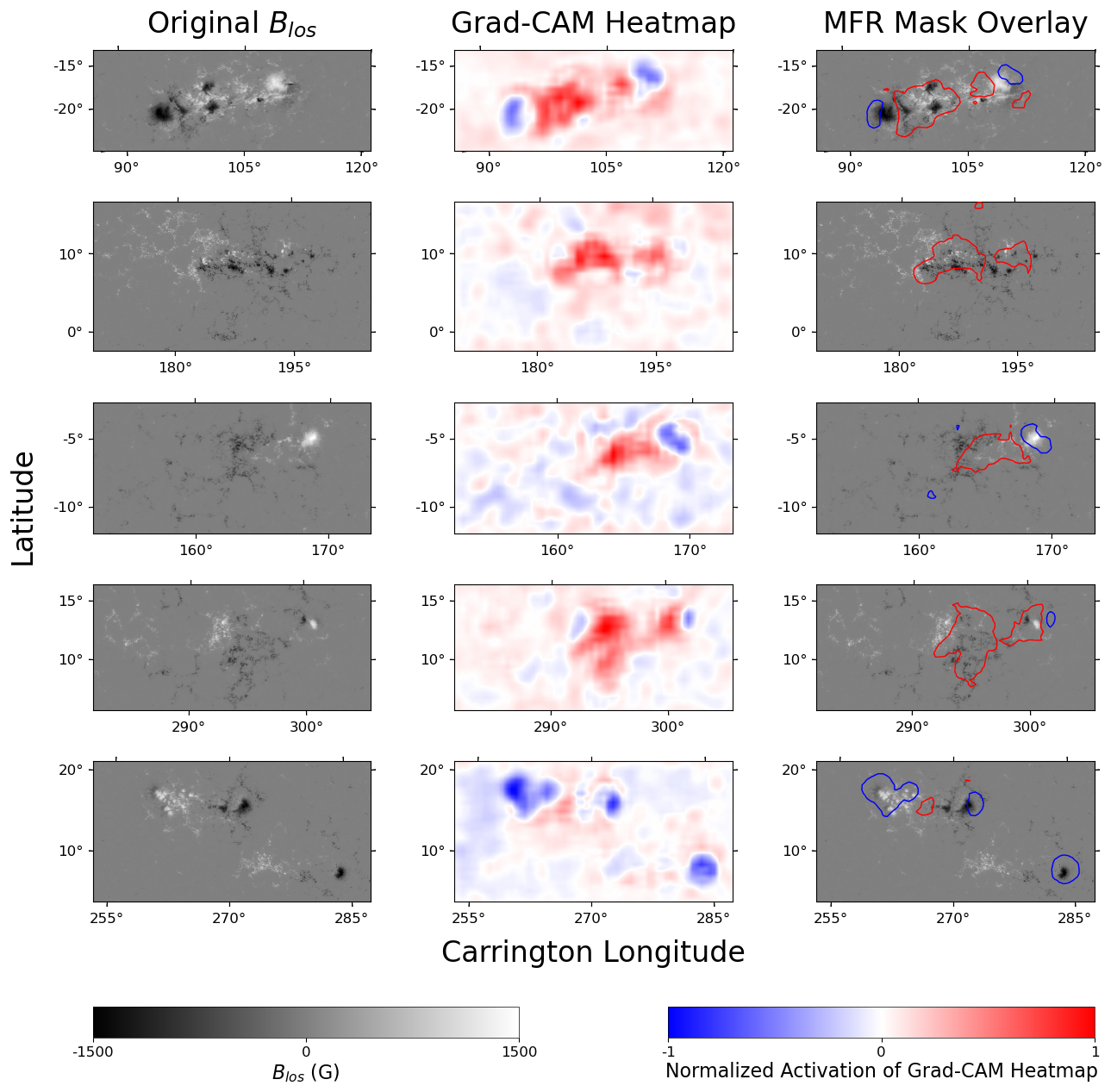}
    \caption{Examples of active regions. (a) HARP 5983 on 30 September 2015 at 03:00:00 TAI. (b) HARP 2693 on 2 May 2013 at 08:00:00 TAI. (c) HARP 6167 on 10 December 2015 at 20:00:00 TAI. (d) HARP 2748 on 22 May 2013 at 08:00:00 TAI. (e) HARP 504 on 16 April 2011 at 08:00:00 TAI. From left to right: Magnetogram ($B_{\mathrm{los}}$), Grad-CAM heatmap (normalized), and MFR mask contour with red (positive) and blue (negative) activations.}
    \label{fig_samples}
\end{figure*}

Analyzing the positive and negative activation regions in Grad-CAM heatmaps is a novel approach to studying MFRs and assessing the complexity of solar active regions. These activations reflect the model's attribution: positive values indicate regions that increase the likelihood of a flare prediction, and negative values indicate regions that decrease the likelihood. Figure~\ref{fig_samples} shows six representative active region examples spanning true positive (a and b), false negative (c), false positive (d), and true negative cases (e). Positive activation areas tend to exhibit mixed polarities, whereas negative areas often correspond to more unipolar or diffuse magnetic structures. This illustrates how different magnetic configurations lead to distinct prediction outcomes.

In the previous section, we performed a parametric model analysis to quantitatively compute sixteen magnetic field parameters from the MFRs. These parameters reflect assessments of magnetic field complexity. However, calculating these parameters requires measurements of vector magnetic fields and complex computations. To derive more intuitive, quantitative features directly from observational data, especially from the Line-of-Sight (LOS) magnetic field, with which to judge magnetic field complexity, we introduce the Polarity Imbalance Index (PII), which is derived from Grad-CAM heatmaps: 
\begin{equation}
  \mathrm{PII} = \frac{|N_p - N_n|}{N_p + N_n}, 
\end{equation}
where $N_p$ and $N_n$ denote the number of pixels exceeding $\pm$150 G in the positive and negative polarities, respectively. We use pixel counts instead of magnetic flux to align with the CNN’s image-based nature. The PII ranges from 0 to 1: the lower values indicate complex and mixed-polarity regions, while the higher values imply the unipolar dominance. The MFR mask was obtained by applying OTSU thresholding to the absolute Grad-CAM heatmap. Within the MFR, pixels were separated into regions of positive and negative activation according to the sign of the Grad-CAM response. For each activation type, the PII was computed using both magnetic polarities in $B_{\mathrm{los}}$ for each connected component within the activation region. Then, it was aggregated using area-weighted averaging. This process quantified the polarity imbalance associated with positive or negative CNN contributions.

\begin{figure*}[ht!]
    \centering
    \includegraphics[width=0.6\linewidth]{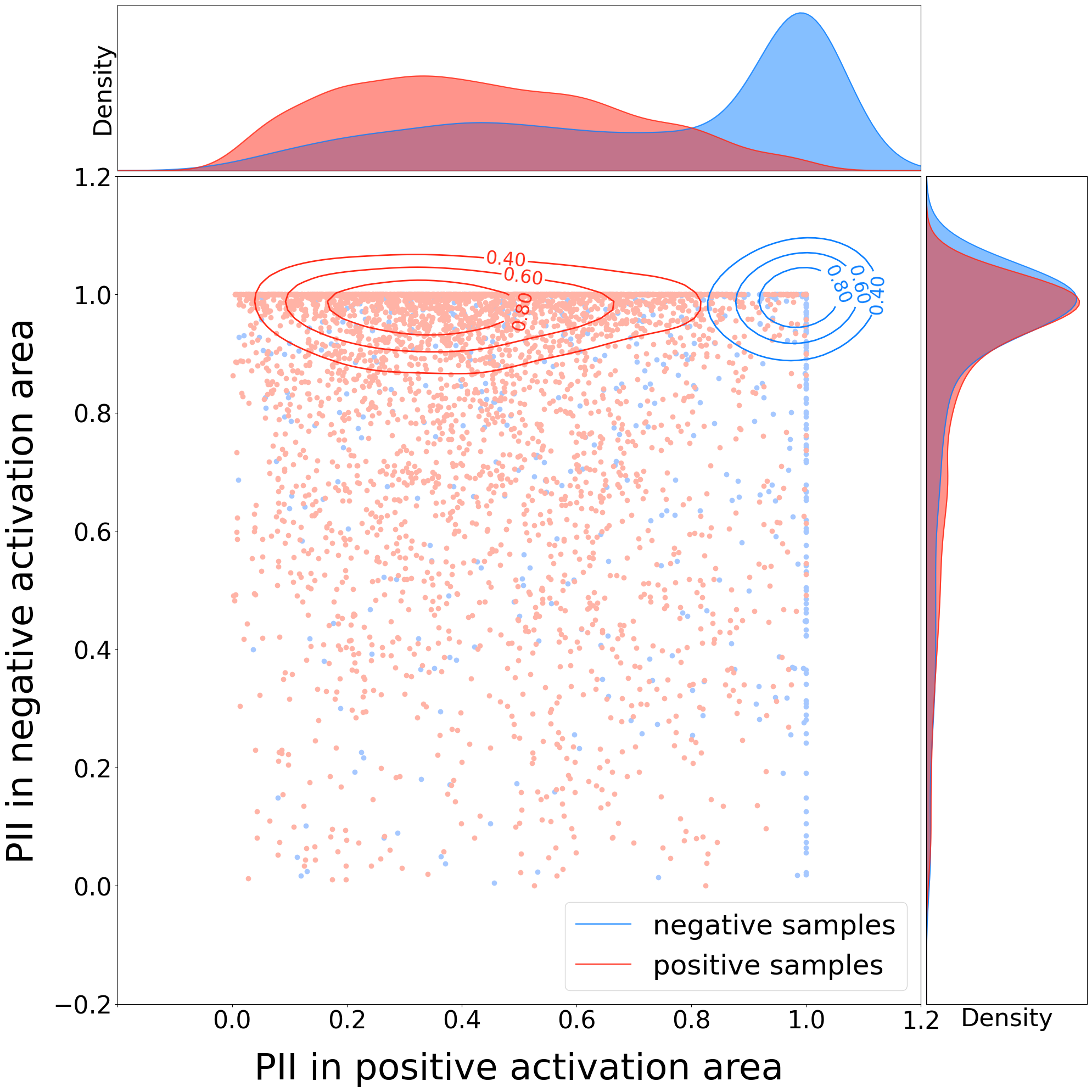}
    \caption{The joint (center) and univariate (top and right) distributions of the Polarity Imbalance Index in positive and negative activation areas across different solar active regions. Each dot represents an active region event. The contours represent probability density: red indicates a positive sample and blue indicates a negative sample. The contours are darker or lighter depending on the percentage value.
    \label{fig_kde}}
\end{figure*}

We use the kernel density estimation (KDE, \citealt{Rosenblatt1956,Parzen1962}) to visualize and analyze the distributional patterns of PII. Figure~\ref{fig_kde} presents the distribution of the PII for positive and negative activation areas in the events that exhibit both. The figure shows the joint (center) and univariate (top and right) distributions of PII in positive and negative activation areas across different solar active region events. Each dot represents an active region event. The contours represent probability density: red indicates a positive sample and blue indicates a negative sample. The contours are darker or lighter depending on the percentage value.

Generally, most of the positive events are concentrated at the top of the KDE map. The PII of the positive activation areas (PII$_p$) with the density peaks around $0.33$, while the PII of the negative activation areas (PII$_n$) peaks around 1. This suggests that the model's positive activation areas tend to correspond to magnetically complex areas, whereas negative activation areas tend to be unipolar. Interestingly, the ratio of positive to negative magnetic polarity regions within active regions prone to solar flares is neither uniform nor unipolar. Rather, one polarity usually dominates. For example, the density peak location of PII$_p$ exhibits a ratio of approximately 1:2. These active regions have sufficient positive or negative magnetic energy and exhibit relatively complex magnetic field configurations with more pronounced PILs, as seen in Figures~\ref{fig_samples}a and b. This distribution suggests that the CNN has learned to associate complex polarity configurations with an increased likelihood of flare occurrence.

In the negative events, both PII$_p$ and PII$_n$ typically have higher mean values and distinct distribution patterns, clustering in the top right corner. The tendency of PII$_n$ to approach 1 in the negative events is particularly evident, and these regions predominantly exhibit single-polarity characteristics. This is consistent with the findings in the positive events. The clustering of PII$_p$ near 1 likely arises from two scenarios. First, negative events may possess both polarities simultaneously. However, the magnetic field is often more diffuse and the positively activated regions are dominated by a single polarity. Second, the activity regions may be unipolar, resulting in high PII values in both activation areas. However, this could be due to the active region being misidentified. These observations suggest that the complexity of the magnetic field within MFRs may be crucial for distinguishing flare-productive active regions from quiescent ones, especially in the areas of positive activation.

\section{Discussion and Summary}\label{sec_summary and discussion}

This study investigates the interpretability and the underlying physical mechanisms of solar flare prediction models based on CNNs through a quantitative magnetic field analysis of MFRs. Unlike previous studies, which relied primarily on qualitative visualizations, we use the Grad-CAM attribution method to extract contiguous MFRs. We then conduct a systematic quantitative validation using sixteen magnetic parameters and a newly introduced PII. Evaluating the predictive capability of the MFR-derived magnetic features reveals a strong physical connection between MFRs and flare occurrence. Based on this result, we further analyze the magnetic complexity of these regions using the PII. Our analysis reveals that the flare-productive active regions are characterized by an intermediately imbalanced distribution of magnetic polarities. These regions are not uniformly distributed or purely unipolar. Rather, they are typically dominated by one polarity. These findings are consistent with the established physical understanding that magnetically complex regions are more prone to flare activity, while regions with simpler magnetic configurations are less likely to produce flares \citep{georgoulis2007magnetic, Toriumi2019}. This physical consistency validates the scientific credibility of the data-driven models and demonstrates that deep learning can implicitly learn the patterns that are consistent with solar physics theory.

The Grad-CAM algorithm used in this study provides region-based attention maps, which offer clear advantages over other interpretability methods that often produce fragmented or sparse visualizations. The resulting contiguous MFRs accurately reflect the physical intuition of spatially localized magnetic activity. Using these MFRs further shows that both active-region localization and the choice of magnetic parameter extraction regions are critical factors in flare prediction. For example, the models based on SHARP and MFR parameters demostrate strong predictive performance, whereas the PIL-based parameters underperform due to the limited spatial extent of the PIL masks (see Figure~\ref{fig_masks}), which leads to substantial information loss. Figure~\ref{fig_masks} compares different types of masks and illustrates how CNN-based methods (via Grad-CAM) and traditional physics-based approaches (SHARP and PIL) localize flare-related magnetic activity. Grad-CAM’s adaptability and sensitivity to magnetic complexity explain its superior predictive performance, thereby further validating the model’s physical reliability. This comparison highlights the ability of deep learning models to capture meaningful physically features while also exposing the limitations of rigid, rule-based masks. Nevertheless, the PIL-based features remain valuable in other applications, such as distinguishing between flare-productive and quiet active regions \citep{wang2019parameters}, classifying flares \citep{li2025prediction}, or hybrid approaches combing PIL and magnetic features \citep{sadykov2017character}.

Beyond the intrinsic “black-box” nature of deep learning models, the choice of mask regions and data processing strategies plays a critical role in the practical deployment of flare prediction systems. A comparative analysis of the predictive capability of magnetic parameters derived from different masks indicates that the spatial extent of the mask significantly impacts model performance. Although the PIL regions are the central sites of magnetic energy release, their limited spatial coverage leads to severe information loss and degraded prediction performance. In contrast, MFR and SHARP masks emphasize different aspects of magnetic activity yet yield comparable predictive ability. These results suggest that combining physical knowledge with XAI techniques can effectively guide the identification of key analysis regions and improve the prediction performance.

Data processing strategies such as resizing the original spatial dimensions to a fixed input size may influence the model performance by altering the scale of the effective pixels in the CNN input. In our preprocessing pipeline, we first zero-pad the four magnetic field components of each SHARP sample to create a square matrix, which we then resize to 512 $\times$ 512 pixels. This approach preserves the complete context and aspect ratio of the active region by normalizing all patches to a uniform input size rather than enforcing a consistent physical pixel scale. However, we acknowledge that variations in the effective pixel scale may affect the learned spatial features and introduce uncertainty. A systematic evaluation of scale-preserving strategies, such as resampling to a uniform physical scale or incorporating scale metadata, is beyond the scope of this study, yet it is a critical direction for future research.

In addition, data-splitting strategies may impact model generalization. We propose a two-stage data-splitting strategy that reveals a critical phenomenon: when the training and testing sets share the same active regions (i.e., information leakage), the model achieves a TSS of 0.762 on Test2, whereas performance is 0.734 on Test1, which consists of entirely independent active regions (see Table~\ref{tab_cnnperform}). This difference indicates that the model trained on historical data may experience performance degradation when applied to new active regions, highlighting the importance of carefully assessing generalization limits in operational settings. Another noteworthy observation is that when transitioning from two-alignment configurations (MFR and SHARP) to three-alignment configurations that include PIL, negative events containing PILs become more difficult to classify, leading to reductions in specificity and TSS (see Table~\ref{appendix_tab_paramperform}). This reflects the impact of the intrinsic sample complexity on real-world model performance.

In summary, this study demonstrates that despite their black-box nature, CNN models can learn physically meaningful representations when trained on large, well-structured datasets. Combining these models with the domain-aware interpretability methods provides predictive capability and serves as an effective tool for physical discovery. Our findings support the idea that integrating data-driven insights with traditional physical knowledge allows XAI to validate known physical mechanisms and facilitate the discovery of new processes that govern solar eruptive events. This advances a data–model integrated paradigm in solar physics research.

\section*{Data Availability}
The data and code are publicly available at \url{https://github.com/ZZsolar/XAI-for-Solar-Flare-Prediction.git}.

\begin{acknowledgments}
We are grateful to the SDO teams for providing the observational data. This work was supported by NSFC under grants 12173019, 12333009, 12127901, the CNSA project D050101, the Fundamental Research Funds for the Central Universities KG202506, and the Young Data Scientist Program of the China National Astronomical Data Center, as well as the AI \& AI for Science Project of Nanjing University.
\end{acknowledgments}

\bibliographystyle{aasjournalv7}
\bibliography{xaisfp}

@article{raissi2019physics,
       author = {{Raissi}, M. and {Perdikaris}, P. and {Karniadakis}, G.~E.},
        title = "{Physics-informed neural networks: A deep learning framework for solving forward and inverse problems involving nonlinear partial differential equations}",
      journal = {Journal of Computational Physics},
         year = 2019,
        month = feb,
       volume = {378},
        pages = {686-707},
          doi = {10.1016/j.jcp.2018.10.045},
       adsurl = {https://ui.adsabs.harvard.edu/abs/2019JCoPh.378..686R}
}

@inproceedings{costa2024leveraging,
title = {Leveraging Physics-Informed Neural Networks as Solar Wind Forecasting Models},
author = {Costa, Nuno and S. Barros, Filipa and Lima, J. and Pinto, Rui and Restivo, André},
booktitle = {Proceedings of the 32nd European Symposium on Artificial Neural Networks, Computational Intelligence and Machine Learning (ESANN 2024)},
year = {2024},
month = {01},
pages = {425-430},
doi = {10.14428/esann/2024.ES2024-110}
}

@article{bobra2015solar,
       author = {{Bobra}, M.~G. and {Couvidat}, S.},
        title = "{Solar Flare Prediction Using SDO/HMI Vector Magnetic Field Data with a Machine-learning Algorithm}",
      journal = {\apj},
         year = 2015,
        month = jan,
       volume = {798},
       number = {2},
          eid = {135},
        pages = {135},
          doi = {10.1088/0004-637X/798/2/135},
archivePrefix = {arXiv},
       eprint = {1411.1405},
 primaryClass = {astro-ph.SR},
       adsurl = {https://ui.adsabs.harvard.edu/abs/2015ApJ...798..135B}
}

@article{liu2017predicting,
   title={Predicting Solar Flares Using SDO/HMI Vector Magnetic Data Products and the Random Forest Algorithm},
   volume={843},
   ISSN={1538-4357},
   url={http://dx.doi.org/10.3847/1538-4357/aa789b},
   DOI={10.3847/1538-4357/aa789b},
   number={2},
   journal={The Astrophysical Journal},
   publisher={American Astronomical Society},
   author={Liu, Chang and Deng, Na and Wang, Jason T. L. and Wang, Haimin},
   year={2017},
   month=jul, pages={104} }

@article{asensio2023review,
       author = {{Asensio Ramos}, Andr{\'e}s and {Cheung}, Mark C.~M. and {Chifu}, Iulia and {Gafeira}, Ricardo},
        title = "{Machine learning in solar physics}",
      journal = {Living Reviews in Solar Physics},
         year = 2023,
        month = dec,
       volume = {20},
       number = {1},
          eid = {4},
        pages = {4},
          doi = {10.1007/s41116-023-00038-x},
archivePrefix = {arXiv},
       eprint = {2306.15308},
 primaryClass = {astro-ph.SR},
       adsurl = {https://ui.adsabs.harvard.edu/abs/2023LRSP...20....4A}
}

@article{liu2019predicting,
       author = {{Liu}, Hao and {Liu}, Chang and {Wang}, Jason T.~L. and {Wang}, Haimin},
        title = "{Predicting Solar Flares Using a Long Short-term Memory Network}",
      journal = {\apj},
         year = 2019,
        month = jun,
       volume = {877},
       number = {2},
          eid = {121},
        pages = {121},
          doi = {10.3847/1538-4357/ab1b3c},
archivePrefix = {arXiv},
       eprint = {1905.07095},
 primaryClass = {astro-ph.SR},
       adsurl = {https://ui.adsabs.harvard.edu/abs/2019ApJ...877..121L}
}

@article{abduallah2023operational,
       author = {{Abduallah}, Yasser and {Wang}, Jason T.~L. and {Wang}, Haimin and {Xu}, Yan},
        title = "{Operational prediction of solar flares using a transformer-based framework}",
      journal = {Scientific Reports},
         year = 2023,
        month = aug,
       volume = {13},
          eid = {13665},
        pages = {13665},
          doi = {10.1038/s41598-023-40884-1},
       adsurl = {https://ui.adsabs.harvard.edu/abs/2023NatSR..1313665A}
}

@article{huang2018deep,
       author = {{Huang}, Xin and {Wang}, Huaning and {Xu}, Long and {Liu}, Jinfu and {Li}, Rong and {Dai}, Xinghua},
        title = "{Deep Learning Based Solar Flare Forecasting Model. I. Results for Line-of-sight Magnetograms}",
      journal = {\apj},
         year = 2018,
        month = mar,
       volume = {856},
       number = {1},
          eid = {7},
        pages = {7},
          doi = {10.3847/1538-4357/aaae00},
       adsurl = {https://ui.adsabs.harvard.edu/abs/2018ApJ...856....7H}
}

@article{sun2022solar,
       author = {{Sun}, Pengchao and {Dai}, Wei and {Ding}, Weiqi and {Feng}, Song and {Cui}, Yanmei and {Liang}, Bo and {Dong}, Zeyin and {Yang}, Yunfei},
        title = "{Solar Flare Forecast Using 3D Convolutional Neural Networks}",
      journal = {\apj},
         year = 2022,
        month = dec,
       volume = {941},
       number = {1},
          eid = {1},
        pages = {1},
          doi = {10.3847/1538-4357/ac9e53},
       adsurl = {https://ui.adsabs.harvard.edu/abs/2022ApJ...941....1S}
}

@article{sun2022predicting,
       author = {{Sun}, Zeyu and {Bobra}, Monica G. and {Wang}, Xiantong and {Wang}, Yu and {Sun}, Hu and {Gombosi}, Tamas and {Chen}, Yang and {Hero}, Alfred},
        title = "{Predicting Solar Flares Using CNN and LSTM on Two Solar Cycles of Active Region Data}",
      journal = {\apj},
         year = 2022,
        month = jun,
       volume = {931},
       number = {2},
          eid = {163},
        pages = {163},
          doi = {10.3847/1538-4357/ac64a6},
archivePrefix = {arXiv},
       eprint = {2204.03710},
 primaryClass = {astro-ph.SR},
       adsurl = {https://ui.adsabs.harvard.edu/abs/2022ApJ...931..163S}
}

@article{wang2019parameters,
       author = {{Wang}, Jingjing and {Liu}, Siqing and {Ao}, Xianzhi and {Zhang}, Yuhang and {Wang}, Tieyan and {Liu}, Yang},
        title = "{Parameters Derived from the SDO/HMI Vector Magnetic Field Data: Potential to Improve Machine-learning-based Solar Flare Prediction Models}",
      journal = {\apj},
         year = 2019,
        month = oct,
       volume = {884},
       number = {2},
          eid = {175},
        pages = {175},
          doi = {10.3847/1538-4357/ab441b},
       adsurl = {https://ui.adsabs.harvard.edu/abs/2019ApJ...884..175W}
}

@article{li2025prediction,
       author = {{Li}, Xuebao and {Li}, Xuefeng and {Zheng}, Yanfang and {Li}, Ting and {Yan}, Pengchao and {Ye}, Hongwei and {Zhang}, Shunhuang and {Wang}, Xiaotian and {Lv}, Yongshang and {Huang}, Xusheng},
        title = "{Prediction of Large Solar Flares Based on SHARP and High-energy-density Magnetic Field Parameters}",
      journal = {\apjs},
         year = 2025,
        month = jan,
       volume = {276},
       number = {1},
          eid = {7},
        pages = {7},
          doi = {10.3847/1538-4365/ad8b2a},
archivePrefix = {arXiv},
       eprint = {2410.18562},
 primaryClass = {astro-ph.SR},
       adsurl = {https://ui.adsabs.harvard.edu/abs/2025ApJS..276....7L}
}

@article{ahmed2013solar,
       author = {{Ahmed}, Omar W. and {Qahwaji}, Rami and {Colak}, Tufan and {Higgins}, Paul A. and {Gallagher}, Peter T. and {Bloomfield}, D. Shaun},
        title = "{Solar Flare Prediction Using Advanced Feature Extraction, Machine Learning, and Feature Selection}",
      journal = {\solphys},
         year = 2013,
        month = mar,
       volume = {283},
       number = {1},
        pages = {157-175},
          doi = {10.1007/s11207-011-9896-1},
       adsurl = {https://ui.adsabs.harvard.edu/abs/2013SoPh..283..157A}
}

@article{bobra2014sharp,
       author = {{Bobra}, M.~G. and {Sun}, X. and {Hoeksema}, J.~T. and {Turmon}, M. and {Liu}, Y. and {Hayashi}, K. and {Barnes}, G. and {Leka}, K.~D.},
        title = "{The Helioseismic and Magnetic Imager (HMI) Vector Magnetic Field Pipeline: SHARPs - Space-Weather HMI Active Region Patches}",
      journal = {\solphys},
         year = 2014,
        month = sep,
       volume = {289},
       number = {9},
        pages = {3549-3578},
          doi = {10.1007/s11207-014-0529-3},
archivePrefix = {arXiv},
       eprint = {1404.1879},
 primaryClass = {astro-ph.SR},
       adsurl = {https://ui.adsabs.harvard.edu/abs/2014SoPh..289.3549B}
}

@inproceedings{selvaraju2017grad,
  author={Selvaraju, Ramprasaath R. and Cogswell, Michael and Das, Abhishek and Vedantam, Ramakrishna and Parikh, Devi and Batra, Dhruv},
  booktitle={2017 IEEE International Conference on Computer Vision (ICCV)}, 
  title={Grad-CAM: Visual Explanations from Deep Networks via Gradient-Based Localization}, 
  year={2017},
  volume={},
  number={},
  pages={618-626},
  doi={10.1109/ICCV.2017.74}}

@ARTICLE{otsu1979ieee,
  author={Otsu, Nobuyuki},
  journal={IEEE Transactions on Systems, Man, and Cybernetics}, 
  title={A Threshold Selection Method from Gray-Level Histograms}, 
  year={1979},
  volume={9},
  number={1},
  pages={62-66},
  doi={10.1109/TSMC.1979.4310076}}

@ARTICLE{yi2021solar,
       author = {{Yi}, Kangwoo and {Moon}, Yong-Jae and {Lim}, Daye and {Park}, Eunsu and {Lee}, Harim},
        title = "{Visual Explanation of a Deep Learning Solar Flare Forecast Model and Its Relationship to Physical Parameters}",
      journal = {\apj},
         year = 2021,
        month = mar,
       volume = {910},
       number = {1},
          eid = {8},
        pages = {8},
          doi = {10.3847/1538-4357/abdebe},
       adsurl = {https://ui.adsabs.harvard.edu/abs/2021ApJ...910....8Y}
}

@misc{bobra2021github,
  author       = {Monica G. Bobra and
                  Xudong Sun and
                  Michael J. Turmon},
  title        = {mbobra/SHARPs: SHARPs 0.1.0 (2021-07-23)},
  month        = jul,
  year         = 2021,
  publisher    = {Zenodo},
  version      = {v0.1.0},
  doi          = {10.5281/zenodo.5131292},
  url          = {https://doi.org/10.5281/zenodo.5131292}
}

@inproceedings{he2016deep,
       author = {{He}, Kaiming and {Zhang}, Xiangyu and {Ren}, Shaoqing and {Sun}, Jian},
        title = "{Deep Residual Learning for Image Recognition}",
    booktitle = {2016 IEEE Conference on Computer Vision and Pattern Recognition (CVPR},
         year = 2016,
        month = jun,
          eid = {1},
        pages = {1},
          doi = {10.1109/CVPR.2016.90},
archivePrefix = {arXiv},
       eprint = {1512.03385},
 primaryClass = {cs.CV},
       adsurl = {https://ui.adsabs.harvard.edu/abs/2016cvpr.confE...1H}
}

@article{srivastava2014dropout,
  author  = {Nitish Srivastava and Geoffrey Hinton and Alex Krizhevsky and Ilya Sutskever and Ruslan Salakhutdinov},
  title   = {Dropout: A Simple Way to Prevent Neural Networks from Overfitting},
  journal = {Journal of Machine Learning Research},
  year    = {2014},
  volume  = {15},
  number  = {56},
  pages   = {1929--1958},
  url     = {http://jmlr.org/papers/v15/srivastava14a.html}
}

@misc{lin2014network,
      title={Network In Network}, 
      author={Min Lin and Qiang Chen and Shuicheng Yan},
      year={2014},
      eprint={1312.4400},
      archivePrefix={arXiv},
      primaryClass={cs.NE},
      url={https://arxiv.org/abs/1312.4400}, 
}

@article{panos2023identifying,
       author = {{Panos}, Brandon and {Kleint}, Lucia and {Zbinden}, Jonas},
        title = "{Identifying preflare spectral features using explainable artificial intelligence}",
      journal = {\aap},
         year = 2023,
        month = mar,
       volume = {671},
          eid = {A73},
        pages = {A73},
          doi = {10.1051/0004-6361/202244835},
archivePrefix = {arXiv},
       eprint = {2301.01560},
 primaryClass = {astro-ph.SR},
       adsurl = {https://ui.adsabs.harvard.edu/abs/2023A&A...671A..73P}
}

@ARTICLE{falconer2002solar,
       author = {{Falconer}, D.~A. and {Moore}, R.~L. and {Gary}, G.~A.},
        title = "{Correlation of the Coronal Mass Ejection Productivity of Solar Active Regions with Measures of Their Global Nonpotentiality from Vector Magnetograms: Baseline Results}",
      journal = {\apj},
         year = 2002,
        month = apr,
       volume = {569},
       number = {2},
        pages = {1016-1025},
          doi = {10.1086/339161},
       adsurl = {https://ui.adsabs.harvard.edu/abs/2002ApJ...569.1016F}
}

@ARTICLE{schrijver2007solar,
       author = {{Schrijver}, Carolus J.},
        title = "{A Characteristic Magnetic Field Pattern Associated with All Major Solar Flares and Its Use in Flare Forecasting}",
      journal = {\apjl},
         year = 2007,
        month = feb,
       volume = {655},
       number = {2},
        pages = {L117-L120},
          doi = {10.1086/511857},
       adsurl = {https://ui.adsabs.harvard.edu/abs/2007ApJ...655L.117S}
}

@ARTICLE{breiman2001random,
       author = {{Breiman}, Leo},
        title = "{Random Forests.}",
      journal = {Machine Learning},
         year = 2001,
        month = jan,
       volume = {45},
        pages = {5-32},
          doi = {10.1023/A:1010933404324},
       adsurl = {https://ui.adsabs.harvard.edu/abs/2001MachL..45....5B}
}

@article{cortes1995support,
	title = {Support-vector networks},
	volume = {20},
	issn = {1573-0565},
	url = {https://doi.org/10.1007/BF00994018},
	doi = {10.1007/BF00994018},
	number = {3},
	journal = {Machine Learning},
	author = {Cortes, Corinna and Vapnik, Vladimir},
	month = sep,
	year = {1995},
	pages = {273--297},
}

@ARTICLE{rumelhart1986learning,
       author = {{Rumelhart}, David E. and {Hinton}, Geoffrey E. and {Williams}, Ronald J.},
        title = "{Learning representations by back-propagating errors}",
      journal = {\nat},
         year = 1986,
        month = oct,
       volume = {323},
       number = {6088},
        pages = {533-536},
          doi = {10.1038/323533a0},
       adsurl = {https://ui.adsabs.harvard.edu/abs/1986Natur.323..533R}
}

@ARTICLE{nishizuka2018prediction,
       author = {{Nishizuka}, N. and {Sugiura}, K. and {Kubo}, Y. and {Den}, M. and {Ishii}, M.},
        title = "{Deep Flare Net (DeFN) Model for Solar Flare Prediction}",
      journal = {\apj},
         year = 2018,
        month = may,
       volume = {858},
       number = {2},
          eid = {113},
        pages = {113},
          doi = {10.3847/1538-4357/aab9a7},
archivePrefix = {arXiv},
       eprint = {1805.03421},
 primaryClass = {astro-ph.SR},
       adsurl = {https://ui.adsabs.harvard.edu/abs/2018ApJ...858..113N}
}

@ARTICLE{ran2022solar,
       author = {{Ran}, Hao and {Liu}, Ying D. and {Guo}, Yang and {Wang}, Rui},
        title = "{Relationship between Successive Flares in the Same Active Region and SHARP Parameters}",
      journal = {\apj},
         year = 2022,
        month = sep,
       volume = {937},
       number = {1},
          eid = {43},
        pages = {43},
          doi = {10.3847/1538-4357/ac80fa},
archivePrefix = {arXiv},
       eprint = {2207.07254},
 primaryClass = {astro-ph.SR},
       adsurl = {https://ui.adsabs.harvard.edu/abs/2022ApJ...937...43R}
}

@article{sadykov2017character,
       author = {{Sadykov}, Viacheslav M. and {Kosovichev}, Alexander G.},
        title = "{Relationships between Characteristics of the Line-of-sight Magnetic Field and Solar Flare Forecasts}",
      journal = {\apj},
         year = 2017,
        month = nov,
       volume = {849},
       number = {2},
          eid = {148},
        pages = {148},
          doi = {10.3847/1538-4357/aa9119},
archivePrefix = {arXiv},
       eprint = {1704.03423},
 primaryClass = {astro-ph.SR},
       adsurl = {https://ui.adsabs.harvard.edu/abs/2017ApJ...849..148S}
}

@article{georgoulis2007magnetic,
  author  = {Georgoulis, M.~K.},
  title   = {Magnetic Complexity in Eruptive Solar Active Regions and Associated Eruption Parameters},
  journal = {Geophysical Research Letters},
  volume  = {34},
  number  = {17},
  year    = {2007},
  doi     = {10.1029/2007GL032040}
}

@misc{jarolim2025pinnme,
      title={PINN ME: A Physics-Informed Neural Network Framework for Accurate Milne-Eddington Inversions of Solar Magnetic Fields}, 
      author={Robert Jarolim and Momchil E. Molnar and Benoit Tremblay and Rebecca Centeno and Matthias Rempel},
      year={2025},
      eprint={2502.13924},
      archivePrefix={arXiv},
      primaryClass={astro-ph.SR},
      url={https://arxiv.org/abs/2502.13924}, 
}

@ARTICLE{Shibata2011,
   author = {{Shibata}, K. and {Magara}, T.},
    title = "{Solar Flares: Magnetohydrodynamic Processes}",
  journal = {Living Reviews in Solar Physics},
     year = 2011,
    month = dec,
   volume = 8,
      eid = {6},
    pages = {6},
      doi = {10.12942/lrsp-2011-6},
   adsurl = {http://adsabs.harvard.edu/abs/2011LRSP....8....6S}
}

@ARTICLE{Fletcher2011,
   author = {{Fletcher}, L. and {Dennis}, B.~R. and {Hudson}, H.~S. and {Krucker}, S. and 
	{Phillips}, K. and {Veronig}, A. and {Battaglia}, M. and {Bone}, L. and 
	{Caspi}, A. and {Chen}, Q. and {Gallagher}, P. and {Grigis}, P.~T. and 
	{Ji}, H. and {Liu}, W. and {Milligan}, R.~O. and {Temmer}, M.
	},
    title = "{An Observational Overview of Solar Flares}",
  journal = {\ssr},
 primaryClass = "astro-ph.SR",
     year = 2011,
    month = sep,
   volume = 159,
    pages = {19-106},
      doi = {10.1007/s11214-010-9701-8},
   adsurl = {http://adsabs.harvard.edu/abs/2011SSRv..159...19F}
}

@ARTICLE{Chen2011,
       author = {{Chen}, P.~F.},
        title = "{Coronal Mass Ejections: Models and Their Observational Basis}",
      journal = {Living Reviews in Solar Physics},
         year = 2011,
        month = apr,
       volume = {8},
       number = {1},
          eid = {1},
        pages = {1},
          doi = {10.12942/lrsp-2011-1},
       adsurl = {https://ui.adsabs.harvard.edu/abs/2011LRSP....8....1C}
}

@ARTICLE{Toriumi2019,
       author = {{Toriumi}, Shin and {Wang}, Haimin},
        title = "{Flare-productive active regions}",
      journal = {Living Reviews in Solar Physics},
         year = 2019,
        month = dec,
       volume = {16},
       number = {1},
          eid = {3},
        pages = {3},
          doi = {10.1007/s41116-019-0019-7},
archivePrefix = {arXiv},
       eprint = {1904.12027},
 primaryClass = {astro-ph.SR},
       adsurl = {https://ui.adsabs.harvard.edu/abs/2019LRSP...16....3T}
}

@ARTICLE{Huang2024,
       author = {{Huang}, Xin and {Zhao}, Zhongrui and {Zhong}, Yufeng and {Xu}, Long and {Kors{\'o}s}, Marianna B. and {Erd{\'e}lyi}, R.},
        title = "{Short-term solar eruptive activity prediction models based on machine learning approaches: A review}",
      journal = {Science China Earth Sciences},
         year = 2024,
        month = dec,
       volume = {67},
       number = {12},
        pages = {3727-3764},
          doi = {10.1007/s11430-023-1375-2},
       adsurl = {https://ui.adsabs.harvard.edu/abs/2024ScChD..67.3727H}
}

@ARTICLE{Schrijver2007,
       author = {{Schrijver}, Carolus J.},
        title = "{A Characteristic Magnetic Field Pattern Associated with All Major Solar Flares and Its Use in Flare Forecasting}",
      journal = {\apjl},
         year = 2007,
        month = feb,
       volume = {655},
       number = {2},
        pages = {L117-L120},
          doi = {10.1086/511857},
       adsurl = {https://ui.adsabs.harvard.edu/abs/2007ApJ...655L.117S}
}

@ARTICLE{Kusano2020,
       author = {{Kusano}, Kanya and {Iju}, Tomoya and {Bamba}, Yumi and {Inoue}, Satoshi},
        title = "{A physics-based method that can predict imminent large solar flares}",
      journal = {Science},
         year = 2020,
        month = jul,
       volume = {369},
       number = {6503},
        pages = {587-591},
          doi = {10.1126/science.aaz2511},
       adsurl = {https://ui.adsabs.harvard.edu/abs/2020Sci...369..587K}
}

@ARTICLE{Hearst1998,
  author={Hearst, M.A. and Dumais, S.T. and Osuna, E. and Platt, J. and Scholkopf, B.},
  journal={IEEE Intelligent Systems and their Applications}, 
  title={Support vector machines}, 
  year={1998},
  volume={13},
  number={4},
  pages={18-28},
  doi={10.1109/5254.708428}}

@article{Breiman2001,
  author       = {Leo Breiman},
  title        = {Random Forests},
  journal      = {Mach. Learn.},
  volume       = {45},
  number       = {1},
  pages        = {5--32},
  year         = {2001},
  url          = {https://doi.org/10.1023/A:1010933404324},
  doi          = {10.1023/A:1010933404324},
  bibsource    = {dblp computer science bibliography, https://dblp.org}
}

@article{Parzen1962,
author = {Emanuel Parzen},
title = {{On Estimation of a Probability Density Function and Mode}},
volume = {33},
journal = {The Annals of Mathematical Statistics},
number = {3},
publisher = {Institute of Mathematical Statistics},
pages = {1065 -- 1076},
year = {1962},
doi = {10.1214/aoms/1177704472},
URL = {https://doi.org/10.1214/aoms/1177704472}
}

@article{Rosenblatt1956,
author = {Murray Rosenblatt},
title = {{Remarks on Some Nonparametric Estimates of a Density Function}},
volume = {27},
journal = {The Annals of Mathematical Statistics},
number = {3},
publisher = {Institute of Mathematical Statistics},
pages = {832 -- 837},
year = {1956},
doi = {10.1214/aoms/1177728190},
URL = {https://doi.org/10.1214/aoms/1177728190}
}

@ARTICLE{xu2025prediction,
       author = {{Xu}, Dan and {Sun}, Pengchao and {Feng}, Song and {Liang}, Bo and {Dai}, Wei},
        title = "{Solar Flare Forecasting Using Hybrid Neural Networks}",
      journal = {\apjs},
         year = 2025,
        month = feb,
       volume = {276},
       number = {2},
          eid = {68},
        pages = {68},
          doi = {10.3847/1538-4365/ada281},
       adsurl = {https://ui.adsabs.harvard.edu/abs/2025ApJS..276...68X}
}

@ARTICLE{tang2021prediction,
       author = {{Tang}, Rongxin and {Liao}, Wenti and {Chen}, Zhou and {Zeng}, Xunwen and {Wang}, Jing-song and {Luo}, Bingxian and {Chen}, Yanhong and {Cui}, Yanmei and {Zhou}, Meng and {Deng}, Xiaohua and {Li}, Haimeng and {Yuan}, Kai and {Hong}, Sheng and {Wu}, Zhiping},
        title = "{Solar Flare Prediction Based on the Fusion of Multiple Deep-learning Models}",
      journal = {\apjs},
         year = 2021,
        month = dec,
       volume = {257},
       number = {2},
          eid = {50},
        pages = {50},
          doi = {10.3847/1538-4365/ac249e},
       adsurl = {https://ui.adsabs.harvard.edu/abs/2021ApJS..257...50T}
}

@ARTICLE{Samek2021explaining,
  author={Samek, Wojciech and Montavon, Grégoire and Lapuschkin, Sebastian and Anders, Christopher J. and Müller, Klaus-Robert},
  journal={Proceedings of the IEEE}, 
  title={Explaining Deep Neural Networks and Beyond: A Review of Methods and Applications}, 
  year={2021},
  volume={109},
  number={3},
  pages={247-278},
  doi={10.1109/JPROC.2021.3060483}}

@ARTICLE{lieu2025comprehensive,
       author = {{Lieu}, Maggie},
        title = "{A Comprehensive Guide to Interpretable AI-Powered Discoveries in Astronomy}",
      journal = {Universe},
         year = 2025,
        month = jun,
       volume = {11},
       number = {6},
          eid = {187},
        pages = {187},
          doi = {10.3390/universe11060187},
       adsurl = {https://ui.adsabs.harvard.edu/abs/2025Univ...11..187L}
}

@article{Arrieta2020,
title = {Explainable Artificial Intelligence (XAI): Concepts, taxonomies, opportunities and challenges toward responsible AI},
journal = {Information Fusion},
volume = {58},
pages = {82-115},
year = {2020},
issn = {1566-2535},
doi = {https://doi.org/10.1016/j.inffus.2019.12.012},
url = {https://www.sciencedirect.com/science/article/pii/S1566253519308103},
author = {Alejandro {Barredo Arrieta} and Natalia Díaz-Rodríguez and Javier {Del Ser} and Adrien Bennetot and Siham Tabik and Alberto Barbado and Salvador Garcia and Sergio Gil-Lopez and Daniel Molina and Richard Benjamins and Raja Chatila and Francisco Herrera}
}

@article{Schubert2017,
    author     = {Schubert, Erich and Sander, J\"{o}rg and Ester, Martin and Kriegel, Hans Peter and Xu, Xiaowei},
    title      = {DBSCAN Revisited, Revisited: Why and How You Should (Still) Use DBSCAN},
    year       = {2017},
    issue_date = {September 2017},
    publisher  = {Association for Computing Machinery},
    address    = {New York, NY, USA},
    volume     = {42},
    number     = {3},
    issn       = {0362-5915},
    url        = {https://doi.org/10.1145/3068335},
    doi        = {10.1145/3068335},
    journal    = {ACM Trans. Database Syst.},
    month      = {jul},
    articleno  = {19},
    numpages   = {21},
}

@ARTICLE{sdo2012,
       author = {{Pesnell}, W. Dean and {Thompson}, B.~J. and {Chamberlin}, P.~C.},
        title = "{The Solar Dynamics Observatory (SDO)}",
      journal = {\solphys},
         year = 2012,
        month = jan,
       volume = {275},
       number = {1-2},
        pages = {3-15},
          doi = {10.1007/s11207-011-9841-3},
       adsurl = {https://ui.adsabs.harvard.edu/abs/2012SoPh..275....3P}
}

@ARTICLE{hmi2012,
       author = {{Scherrer}, P.~H. and {Schou}, J. and {Bush}, R.~I. and {Kosovichev}, A.~G. and {Bogart}, R.~S. and {Hoeksema}, J.~T. and {Liu}, Y. and {Duvall}, T.~L. and {Zhao}, J. and {Title}, A.~M. and {Schrijver}, C.~J. and {Tarbell}, T.~D. and {Tomczyk}, S.},
        title = "{The Helioseismic and Magnetic Imager (HMI) Investigation for the Solar Dynamics Observatory (SDO)}",
      journal = {\solphys},
         year = 2012,
        month = jan,
       volume = {275},
       number = {1-2},
        pages = {207-227},
          doi = {10.1007/s11207-011-9834-2},
       adsurl = {https://ui.adsabs.harvard.edu/abs/2012SoPh..275..207S}
}

@ARTICLE{zhang2022,
       author = {{Zhang}, Hewei and {Li}, Qin and {Yang}, Yanxing and {Jing}, Ju and {Wang}, Jason T.~L. and {Wang}, Haimin and {Shang}, Zuofeng},
        title = "{Solar Flare Index Prediction Using SDO/HMI Vector Magnetic Data Products with Statistical and Machine-learning Methods}",
      journal = {\apjs},
         year = 2022,
        month = dec,
       volume = {263},
       number = {2},
          eid = {28},
        pages = {28},
          doi = {10.3847/1538-4365/ac9b17},
archivePrefix = {arXiv},
       eprint = {2209.13779},
 primaryClass = {astro-ph.SR},
       adsurl = {https://ui.adsabs.harvard.edu/abs/2022ApJS..263...28Z}
}

@ARTICLE{Newman2025ApJS,
       author = {{Newman}, Timothy S. and {Hall}, Christian W. and {Farris}, Luke and {Singh}, Talwinder and {Pogorelov}, Nikolai and {Benson}, Bernard and {Raza}, Syed and {Trital}, Prajun},
        title = "{Solar Flare Forecasting Using Machine Learning and SDO/HMI Data: A Multiple Machine Learning Model and Data Curation Technique Comparison Study}",
      journal = {\apjs},
         year = 2025,
        month = oct,
       volume = {280},
       number = {2},
          eid = {54},
        pages = {54},
          doi = {10.3847/1538-4365/adf8e0},
       adsurl = {https://ui.adsabs.harvard.edu/abs/2025ApJS..280...54N}
}

@ARTICLE{Li2025,
       author = {{Li}, Xuebao and {Li}, Xuefeng and {Zheng}, Yanfang and {Li}, Ting and {Yan}, Pengchao and {Ye}, Hongwei and {Zhang}, Shunhuang and {Wang}, Xiaotian and {Lv}, Yongshang and {Huang}, Xusheng},
        title = "{Prediction of Large Solar Flares Based on SHARP and High-energy-density Magnetic Field Parameters}",
      journal = {\apjs},
         year = 2025,
        month = jan,
       volume = {276},
       number = {1},
          eid = {7},
        pages = {7},
          doi = {10.3847/1538-4365/ad8b2a},
archivePrefix = {arXiv},
       eprint = {2410.18562},
 primaryClass = {astro-ph.SR},
       adsurl = {https://ui.adsabs.harvard.edu/abs/2025ApJS..276....7L}
}

@inproceedings{nari2010relu,
author = {Nair, Vinod and Hinton, Geoffrey E.},
title = {Rectified linear units improve restricted boltzmann machines},
year = {2010},
isbn = {9781605589077},
publisher = {Omnipress},
address = {Madison, WI, USA},
booktitle = {Proceedings of the 27th International Conference on International Conference on Machine Learning},
pages = {807–814},
numpages = {8},
location = {Haifa, Israel},
series = {ICML'10}
}

\end{document}